\documentclass{elsart}
\usepackage{graphicx}
\usepackage{txfonts}
\usepackage{amssymb}

\begin{document}
\begin{frontmatter}
\title{Amplitude calibration of a digital radio antenna array for
  measuring cosmic ray air showers}

   \author[inst1]{S.~Nehls\corauthref{cor}},
   \corauth[cor]{Corresponding author.}
   \ead{steffen.nehls@ik.fzk.de}
   \author[inst2]{A.~Hakenjos},
   \author[inst3]{M.J.~Arts},
   \author[inst1,inst2]{J.~Bl\"umer},
   \author[inst1]{H.~Bozdog},
   \author[inst3]{W.A.~van Cappellen},
   \author[inst4,inst3]{H.~Falcke},
   \author[inst1]{A.~Haungs},
   \author[inst4]{A.~Horneffer},
   \author[inst1]{T.~Huege},
   \author[inst1]{P.G.~Isar}, and
   \author[inst5]{O.~Kr\"omer}

   \address[inst1]{Institut f\"ur Kernphysik, Forschungszentrum Karlsruhe,
                   76021 Karlsruhe, Germany}
   \address[inst2]{Institut f\"ur Experimentelle Kernphysik, Universit\"at 
                   Karlsruhe, 76021 Karlsruhe, Germany}
   \address[inst3]{ASTRON, 7990 AA Dwingeloo, The Netherlands}
   \address[inst4]{Dpt. Astrophysics, Radboud University, 6525 ED Nijmegen,
                   The Netherlands}
   \address[inst5]{Institut f\"ur Prozessdatenverarbeitung und Elektronik, 
                   Forschungszentrum Karlsruhe, 76021 Karlsruhe, Germany}
 
  \begin{abstract}
    Radio pulses are emitted during the development of air showers,
    where air showers are generated by ultra-high energy cosmic rays
    entering the Earth's atmosphere. These nano-second short pulses
    are presently investigated by various experiments for the purpose
    of using them as a new detection technique for cosmic particles.
    For an array of 30 digital radio antennas (LOPES experiment) an
    absolute amplitude calibration of the radio antennas including the
    full electronic chain of the data acquisition system is performed,
    in order to estimate absolute values of the electric field
    strength for these short radio pulses. This is mandatory, because
    the measured radio signals in the MHz frequency range have to be
    compared with theoretical estimates and with predictions from
    Monte Carlo simulations to reconstruct features of the primary
    cosmic particle. A commercial reference radio emitter is used to
    estimate frequency dependent correction factors for each single
    antenna of the radio antenna array. The expected received power is
    related to the power recorded by the full electronic chain.
    Systematic uncertainties due to different environmental conditions
    and the described calibration procedure are of order 20~\%.
   \end{abstract}

\end{frontmatter}
%

\section{Introduction}
Since Viktor Hess discovered the existence of cosmic rays in the early
20th century, there have been many different experiments and methods
for the measurement of this radiation. These cosmic rays consist
mainly of ionized atomic nuclei originating mostly from extra-solar
sources with energies from a few $10^9$~eV to above $10^{20}$~eV. The
measurements of these particles are based on two different techniques.
In the energy range up to $10^{14}$~eV direct measurements in space or
with balloons are possible. Above this energy, it is necessary to use
so-called indirect methods. Here, one measures the resulting particle
shower when a cosmic ray nucleus interacts with nuclei of the air
molecules.  The first high-energy interaction at higher altitudes is
followed by a cascade of secondary interactions, which creates an
extended disc of particles.  This shower disc consists of a hadronic
component (approximately 1~\% of the shower particles at sea level),
muons (approximately 10~\%), and an electromagnetic component
(electrons and positrons approximately 90~\%)~\cite{Hau03}.

Beside generating these secondary particles, the interaction of the
electromagnetic component of the shower with the surrounding medium
causes secondary radiation based on different mechanisms.  Excitation
of nitrogen atoms in the atmosphere leads to fluorescence radiation
and ultra-relativistic electrons and positrons emit Cherenkov
radiation, both at a few hundred nanometers wavelength.  For radiation
in the radio frequency range the origin is most likely connected with
the Earth's magnetic field, already suggested in the late 1960's by
Hazen et al.~\cite{Haze69} as geomagnetic production. In the early
1970's experiments measured the radio pulses in EAS and could verify
the order of magnitude for the absolute field strength.  These
experiments were summarized in an excellent review of Allan et
al.~\cite{Alla71}.  Nevertheless there was a lack of understanding of
the theory for the geomagnetic effect, leading to large uncertainties
for the detected radio pulses~\cite{Atra78}.

In 2003 Falcke and Gorham~\cite{Falc03} considered the radiation as a
coherent geosynchrotron effect which is a result of electron-positron
pairs being deflected in the terrestrial magnetic field and emitting
synchrotron radiation. Taking into consideration the overlap of
different pulses and the properties of the shower, the results are
several tens of nanosecond long radio pulses. They can be
measured nearly unattenuated at ground level. Within this framework of
the geosynchrotron effect, values for the electric field strength at
ground level were predicted, first with analytical~\cite{Hueg05b}, and
later with full Monte Carlo calculations~\cite{Hueg07}. By these
simulations it was proposed that the field strength emitted by air
showers with primary energies above $10^{17}\,$eV should be detectable
by an antenna array like the LOPES experiment~\cite{Horn06a}.

The LOPES digital radio antenna field of thirty east-west polarized
short dipole antennas is placed inside the existing multiple
detector-component air shower experiment
KASCADE-Grande~\cite{Anto03,Nava04} and measures the radio pulse
emitted by the particle shower.  But, to finally compare the measured
pulses with expectations from detailed simulations the measured
amplitudes of the antennas have to be calibrated, i.e.~the conversion
factor from ADC counts to electric field strength amplitude per
bandwidth in [$\mathrm{\mu V\,m^{-1}\,MHz^{-1}}$] has to be estimated.

This paper describes a method for an almost complete end-to-end
amplitude calibration of an antenna array like LOPES. The purpose of
LOPES is the measurement of the absolute electric field strength of
radio pulses emitted by extensive air showers.

\section{The LOPES antenna array}
The LOfar PrototypE Station -- LOPES consists of thirty dipole
antennas distributed over the field of the KASCADE-Grande experiment
in the Forschungszentrum Karlsruhe, Germany.  The antennas were
designed as prototypes for the LOFAR experiment~\cite{Rotg06}, a large
antenna array for astronomical purposes.  LOFAR is presently being
built in The Netherlands.  The initial idea for LOPES was to test the
potential of the LOFAR setup for the measurement of cosmic rays and to
investigate the properties of the radio component of an extensive air
shower. A related experiment, CODALEMA~\cite{Ardo05}, at the Nan{\c
  c}ay radio observatory is also investigating this radio component.

\begin{figure}
  \centering
  \includegraphics[width=11.0cm]{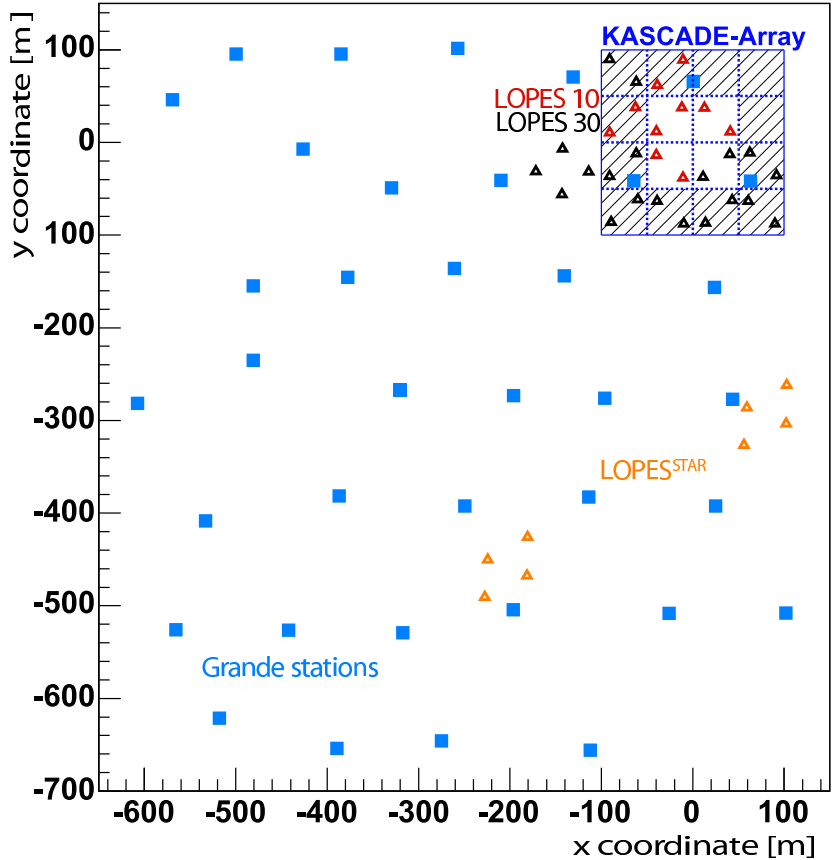}
  \caption{Sketch of the KASCADE-Grande and LOPES 
    experiments: The KASCADE particle detector array and the
    distribution of the 37 stations of the Grande array are shown.
    The location of the 30 LOPES radio antennas as well as the
    LOPES$^{\rm STAR}$ antennas is also displayed.}\label{FigLayout}
\end{figure}

The aim of the LOPES experiment is to correlate the observables of the
radio measurements with the shower properties provided by the particle
air-shower experiment KASCADE-Grande.  For this reason LOPES is
triggered by KASCADE-Grande and uses the reconstructed shower data as
input for the pulse analysis. In other words, the shower core position
at ground and the direction of the shower axis are used as starting
values for the reconstruction of the radio signals. A layout of the
experimental setup is sketched in figure~\ref{FigLayout}.
KASCADE-Grande consists mainly of stations equipped with scintillation
detectors, where 252 stations compose the KASCADE array, and further
37 large stations the Grande array.  LOPES consists of 30 LOFAR-type
antennas as well as newly designed antennas forming the LOPES$^{\rm
  STAR}$ array~\cite{Gemm06}. The main purpose of LOPES$^{\rm STAR}$
is to optimize the hardware for an application of this measuring
technique to large scales, e.g.~at the Pierre Auger
Observatory~\cite{Abra04}. All antennas are optimized to measure in
the relatively noise-free frequency range of 40~MHz to 80~MHz.

\begin{figure}
  \centering 
  \includegraphics[width=11cm]{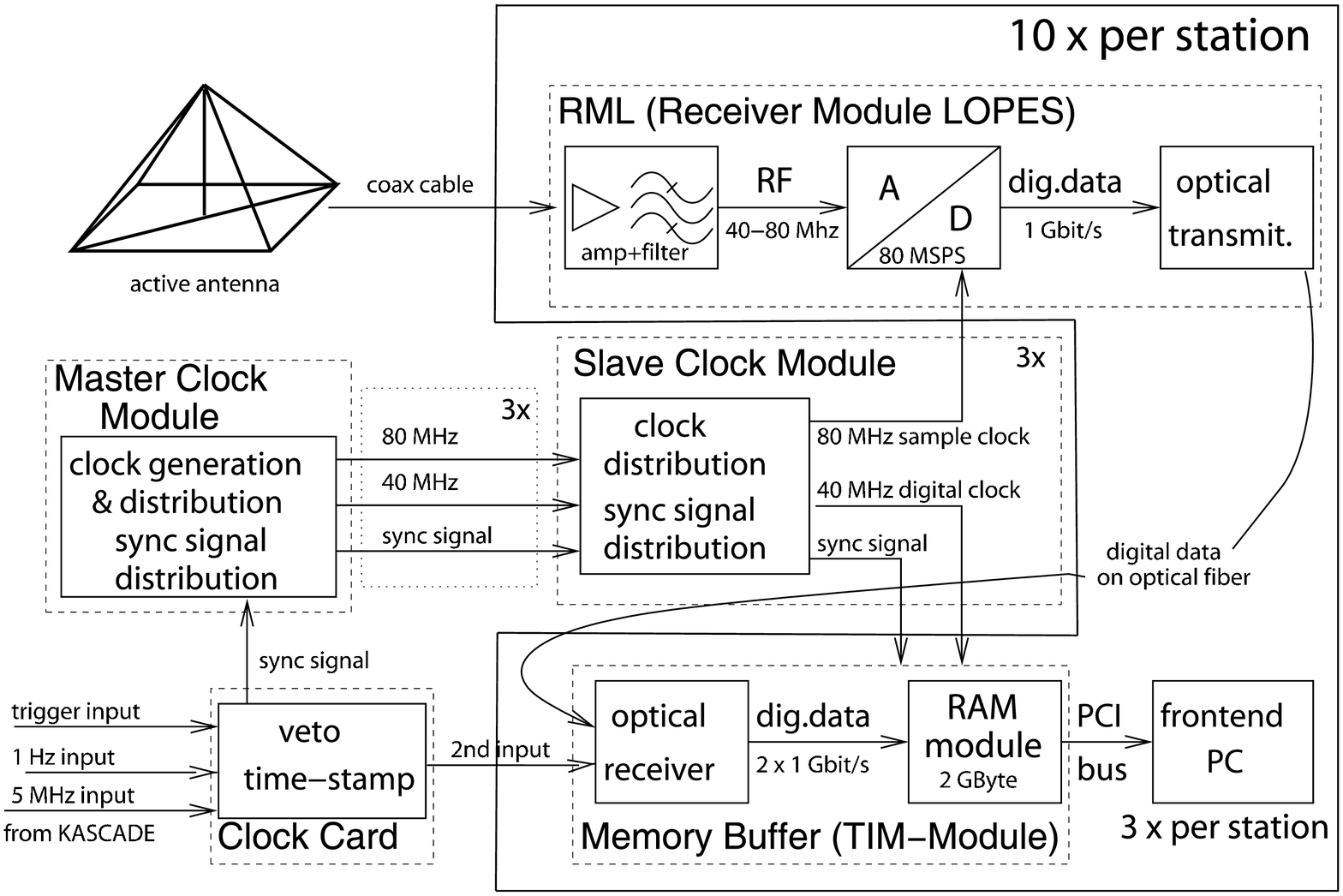}
  \caption{Scheme of LOPES electronics: Incoming radio pulses 
    from EAS are received with the active antenna and transmitted via
    a long coax cable (RG-213) to the Receiver Module (RML).  The
    amplified and bandpass filtered analog signals are digitized with
    a 12-bit A/D converter. After the optical fiber transmission the
    signals are stored in a memory buffer. Trigger from KASCADE-Grande
    are distributed by a master clock module to each station and
    finally transmitted to the front-end PC for data
    read-out.}\label{FigElec}
\end{figure}

The LOPES antennas are inverted-V dipole antennas, where the angle
between the wires is $85^\circ$, and the length of the two arms
correspond to $\lambda/2$ at 75~MHz.  The wires are placed inside
plastic tubes for protection. These tubes are standing atop of a metal
pedestal to protect them from maintenance works at the KASCADE field.
The antennas are connected to low noise amplifiers (LNA). The signal
is then transmitted via a coax cable to one of three stations where it
is again amplified, filtered and digitized. The three stations with 10
antennas each are operated by independent electronic chains and data
acquisition systems. One of these three stations was forming the
initial LOPES10 experiment~\cite{Falc05}. The data is ring-buffered
for $6.25$ seconds and read-out only if an external trigger arrives
(from KASCADE-Grande). Then the central DAQ-PC collects $0.82$
milliseconds of data around the trigger from all 30 antennas, adding a
KASCADE-Grande time-stamp, and stores them as one LOPES-event file.
Figure~\ref{FigElec} displays the scheme of the LOPES electronics in
more detail.
\begin{figure}
  \centering
  \includegraphics[width=8cm]{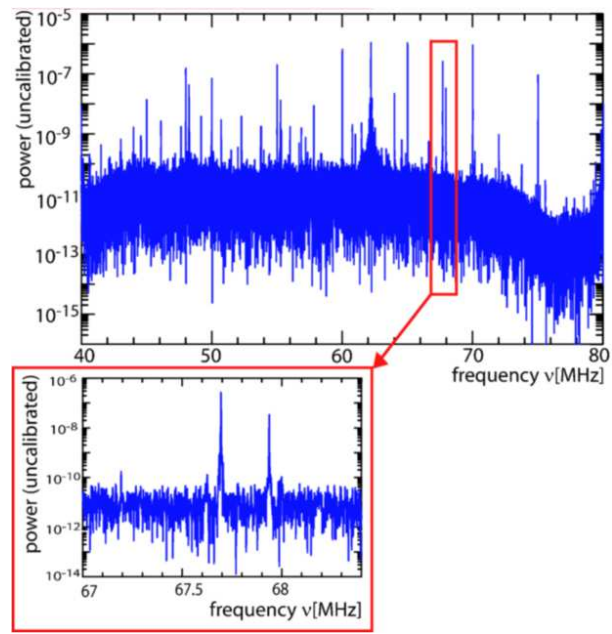}
  \caption{Typical, uncalibrated LOPES raw signal in the 
    frequency domain. The zoom-in shows two audio signal carriers from
    a TV transmitter at 67.70~MHz and 67.92~MHz in more
    detail.}\label{FigTV}
\end{figure}

In figure~\ref{FigTV} an uncalibrated power spectrum from a
LOPES-event file is shown.  The narrow band radio emitters exceed the
noise floor by orders of magnitude. During the analysis procedure
these narrow band radio frequency interferences (RFI), like from a
TV-transmitter as shown in the zoom-in of figure~\ref{FigTV}, are
removed.  With a RFI mitigation it is possible to strongly suppress
most of these narrow emitters, which leads to a better signal-to-noise
ratio for the detected radio pulses.

Finally, a beam-forming procedure is applied to search for coherent
peaks coming from a specific direction. This digital beam-forming
consists of a time shift of the data according to the given direction
and afterwards a combination of the data to calculate the resulting
beam from all antennas.  At this point, the reconstructed shower
direction from KASCADE-Grande is used as a starting value.  For
details of the experimental setup of the LOPES radio antennas see
references~\cite{Horn06a,Falc05,Horn06b}.

First analysis results of the measurements with the LOPES10 setup have
already been published~\cite{Falc05,Apel06,Buij07,Petr07}.  All the
results published so far, however, rely on relative field strength
values, which were only roughly normalized by a fudge factor.  To
quantify the measurements and to compare with theoretical expectations
the antennas need the absolute amplitude calibration.

\section{Amplitude Calibration}
Compared to the smaller initial LOPES10 setup with only one
DAQ-station, for the LOPES30 setup larger variations in the behavior
of the individual electronic channels can be expected. Therefore, a
relative measure of the electric field strengths, as it was done for
the initial LOPES10 setup, is not sufficient. Instead, an amplitude
calibration has to be performed for the whole setup, including the
full electronic chain of the antenna, i.e. LNA, cable, filter, and
analog-to-digital converter.

In general, to perform an amplitude calibration of the antennas there
are two possibilities: either the antenna characteristics are
calculated theoretically and the electronic chain is calibrated in the
laboratory, or the full chain is calibrated by an external source of
known field strength. The former method was performed for the LOPES10
configuration and has only applicability for relative comparisons of
radio emission and air shower properties. The latter procedure has the
advantage of calibrating the full chain of the experiment, the antenna
and the electronics simultaneously (end-to-end calibration). In
addition, this approach provides the opportunity to investigate
influences of environmental variables such as weather or ground
humidity during shower measurements and to test the simulated antenna
gain pattern.

For LOPES30 we have chosen a combination of end-to-end calibration
with the determination of the behavior of individual components. This
hybrid method requires an antenna gain simulation. A full experimental
test of these simulation results requires dedicated measurements, which
are partially performed for this analysis. As external source for a
calibration one can either use an astronomical one, like the galactic
background radiation~\cite{Dulk01}, or a man-made radio source.  For
the location of the LOPES-experiment the galactic background radiation
is fully oblique due to man-made noise sources at Forschungszentrum
Karlsruhe and can therefore not be used as reference source.  What
follows is the description how LOPES30 is amplitude calibrated with
the help of an external reference source.

\subsection{Method}
The amplitude calibration is based on two ingredients: First,
in each individual antenna the signal received from a calibrated radio
source is measured. Second, the expected signal to be
received is calculated using antenna characteristics obtained from
simulations. Then, we compare the expected signal strength with the
measured signal and derive a amplification factor as the ratio of
measured to expected signal, describing how the system alters the
incoming signals.

The main characteristic of the antenna is defined by its directivity.
This directivity describes the reception or emission features for each
direction in relation to a mathematical reference antenna like the
ideal isotropic radiator. The ratio between the power received by a
real antenna and the ideal isotropic radiator for a given direction is
called antenna gain $G$. This antenna gain can be obtained by
measurements or simulations. Experience within the LOFAR initial test
station (ITS~\cite{Nigl07}) has shown that the measurements
contain larger uncertainties than the simulations. Therefore, the
calibration is mainly realized for the zenith direction. For
other directions we will rely basically on simulations. Nevertheless,
some test measurements are performed to verify the simulated values of
the full directivity as the method allows to measure the full
direction characteristics and to evaluate the antenna gain 
simulation (section~\ref{subsec:Direct}).

The method is based on the fact, that the frequency dependent electric
field strength $E_\mathrm{t}(\nu)$ at a certain distance is known.
This electric field strength can also be expressed as known output
power $G_\mathrm{t}P_\mathrm{t}(\nu)$ of the commercial radio source.
The commercially calibrated radio source is used as emitter which LOPES
measures within a calibration mode from artificially triggered events.
The simulated antenna gain $G_\mathrm{r}$ of the LOPES antennas is an
input value for the calculation of the power which is received by the
LOPES electronics.  This calculated input power $P_\mathrm{R}(\nu)$ is
compared with the measured power $P_\mathrm{M}(\nu)$.  The result is a
frequency dependent amplification factor $V(\nu)$ which will be used
during the analysis of the LOPES air shower data to calculate the 
absolute field strength:
\begin{equation}\label{Formel_V}
V(\nu) =\frac{P_\mathrm{M}(\nu)}{P_\mathrm{R}(\nu)}
       =\left(\frac{4\pi r \nu}{c}\right)^2 
        \frac{P_\mathrm{M}(\nu)}{G_\mathrm{r}(\theta,\phi,\nu) G_\mathrm{t}
              P_\mathrm{t}(\nu)\cos^2(\beta)}
\end{equation}
with
\begin{description}
\item[$P_\mathrm{M}$] The power measured with the LOPES antenna and
  calculated in the frequency domain.
\item[$P_\mathrm{R}$] The (calculated) incoming power to the LOPES
  electronics chain.
\item[$\nu$] Frequency of the emitted signal.
\item[$r$] Distance between the external reference source and the
  LOPES antenna.
\item[$G_\mathrm{r}(\theta,\phi,\nu)$] Gain of the LOPES antenna taken
  from simulations.
\item[$G_\mathrm{t}P_\mathrm{t}(\nu)$] Product of the reference source
  antenna gain $G_\mathrm{t}$ and its power $P_\mathrm{t}$ which are
  not exactly known by themselves, but known as the electric field
  strength $E_\mathrm{t}(\nu)$ in $r=10$~m distance. This value is
  obtained from the manufacture calibration report of our reference source.
\item[$\beta$] Angle between the polarization axis of the reference
  source and the field antenna. This axis has to be aligned during
  the measurements.
\end{description}

\subsection{\label{subsec:Simu}Simulation of the antenna gain}
The simulations were performed with the program package IE3D from
Zeland company (see also www.zeland.com), a program using the
multi-pole expansion to calculate the electromagnetic properties of
three-dimensional antennas.  As input to the simulations the real
geometry of the LOPES antenna including the metal pedestal is used.
The inverted-V dipole of the LOPES antenna has an opening angle of
$\alpha\approx 85^\circ$ which alters the directivity pattern compared
to a linear dipole. Taking into account that at 75\,MHz it represents
a $\lambda/2$ dipole the directivity pattern is determined.  For the
definition of the ground an infinite plane is assumed and the
electrical properties of this ground plane are similar to the real
ground. Therefore resistance, ground humidity, and conductivity are
chosen to be close to the actual condition at the antenna field. For
the dielectric constant the value was set to $\epsilon_r=3$ and a
value of $\sigma=0.01$~S~m$^{-1}$ was used for the conductivity. The
simulation describes only the antenna itself, and therefore no antenna
coupling was considered. Hence, a possible effect of mismatch can not
be seen directly from the simulations, but will be taken into
consideration automatically by our method of end-to-end amplitude
calibration.  The simulation ranges from 10~MHz to 100~MHz and
results in a gain value for each frequency (1~MHz step size) and
direction (zenith angle $\theta$ with $5^\circ$ and azimuth angle
$\phi$ with $10^\circ$ step size) of the incoming signal, which can be
displayed as a directional diagram.
\begin{figure}
  \centering 
  \includegraphics[width=8cm]{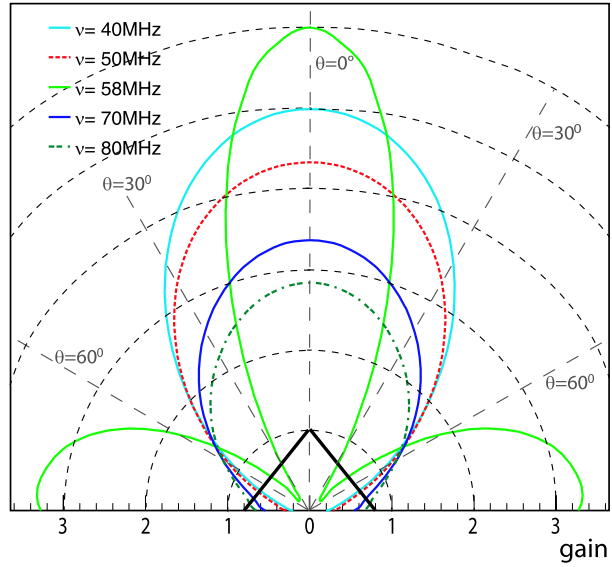}
  \caption{Antenna gain $G_\mathrm{r}(\theta,180^\circ,\nu)$ of the LOPES 
    antenna in the $\vec{E}$-plane (zenithal sensitivity). The antenna
    illustrated by a triangle in the center of the x-axis lies in the
    image plane.}\label{FigSimuXZ}
\end{figure}
\begin{figure}
  \centering 
  \includegraphics[width=7cm]{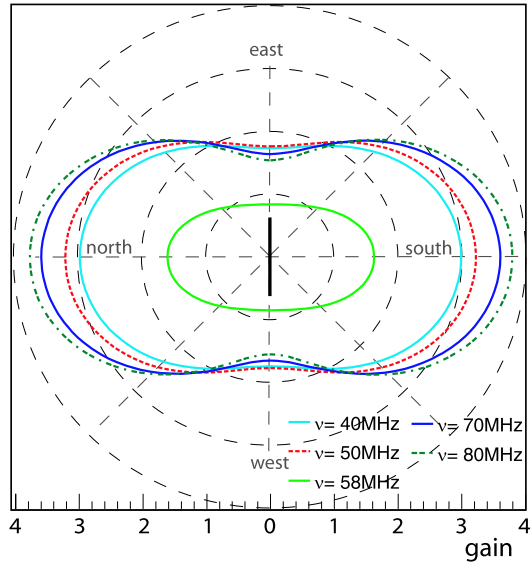}
  \caption{Antenna gain $G_\mathrm{r}(45^\circ,\phi,\nu)$ of the LOPES 
    antenna for a fixed zenith angle (azimuthal sensitivity).  This
    can be considered as a horizontal plane cut at $\theta=45^\circ$
    in figure \ref{FigSimuXZ}. The dipole antenna lies on the
    y-axis.}\label{FigSimuXY}
\end{figure}

In figure~\ref{FigSimuXZ}, a vertical cut through the simulated
diagram is shown and the antenna gain
$G_\mathrm{r}(\theta,180^\circ,\nu)$ is displayed for different
frequencies. The gain is expressed as a relative value to the
sensitivity of an isotropic radiator. The main lobe of the antenna has
$-3$~dB angle of $\approx 70^\circ$. Side lobes are strongly
suppressed due to the pedestal of the LOPES antenna and the ground
plane conditions. This $-3$~dB angle is the full width half maximum
(FWHM) angular beam width. For zenith angles $\theta$ above $60^\circ$
the gain drops significantly compared to the line of symmetry (in the
zenith, $\theta=0^\circ$).

In figure~\ref{FigSimuXY}, a horizontal cut through the simulated
diagram at a fixed zenith angle of $\theta=45^\circ$ is shown. The
visible main lobes at such a zenith angle display an oval shape and
the dipole origin is only indicated by the constriction along the y-axis
at gain zero on the x-axis.  It is obvious that the antennas prefer
radio pulses from north or south as a result of its orientation
(i.e.~here the east-west direction).

An interesting feature is the behavior at 58~MHz, visible as
horizontal side lobes in figure~\ref{FigSimuXZ}. The simulations
indicate that there is a resonance induced from the 2~m by 2~m metal
pedestal below the antenna. The resonance seems to be sharp and
should be visible in the calibration procedure but will be checked
by specific measurements (section~\ref{subsec:Direct}).

\subsection{\label{subsec:VSQ}The reference antenna}
The discussed calibration method relies on an external radio source
which is calibrated independently. For the actual amplitude
calibration method a commercial product from the company Schaffner,
Augsburg (Type: VSQ~1000 with DPA~4000 and RSG~1000) was used (see
also data sheets at www.teseq.com or reference~\cite{Hake06}).

\begin{figure}
  \centering \includegraphics[width=7cm]{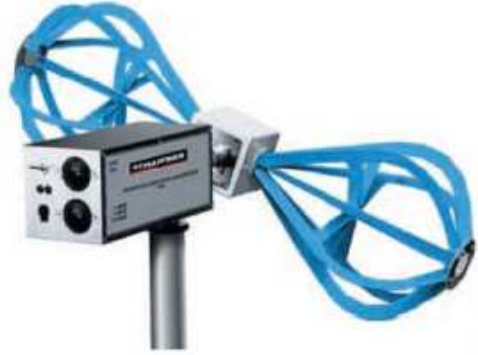}
  \caption{The VSQ~1000 from Schaffner consists of the signal generator
    RSG~1000 and the biconical antenna DPA~4000. The total length of
    the biconical antenna is $40.5\,$cm (image courtesy of
    Schaffner).}\label{FigSchaffner}
\end{figure}

The setup (VSQ), see fig.~\ref{FigSchaffner}, consists of a biconical
antenna, Schaffner DPA~4000, attached to a signal comb-generator
RSG~1000 which gives a signal peak at multiples of 1~MHz from 1~MHz to
1~GHz.  The signal generator has a mean power of $1\,\mu$W over the
whole LOPES frequency range (40--80~MHz). Since it is
battery-operated, it is usable for measurements on the antenna field.
The biconical antenna DPA~4000 is linearly polarized and has a nearly
constant directivity close to its main lobe.  This is important since
it results in only a small loss should the radio source be slightly
off target.  The DPA~4000 is originally designed for the frequency
range 300--1000~MHz, but the VSQ~1000 is specified and certificated
for the broader frequency range from 30--1000~MHz despite the fact
that the antenna factor changes in the lower frequency range.
Nevertheless, it fits within the LOPES frequency range.

The reference radio source itself is calibrated, hence the resulting
field strength at any given distance can be calculated. For the
fiducial distance of 10~m from the VSQ the electric field strength
ranges from $210~\mathrm{\mu V\,m^{-1}\,MHz^{-1}}$ at 40~MHz up to
$2100~\mathrm{\mu V\,m^{-1}\,MHz^{-1}}$ at 80~MHz. The systematic
uncertainty of the calibration measurements for the electric field
strength is $2.5$~dB (or $\approx 30$~\%). This value is reported in
the certificate of calibration provided by the manufacturer for our
individual VSQ. Beside this uncertainty of the calibration method the
signal stability of the RSG~1000 for the temperature range $10^\circ$C
--$30^\circ$C is $<0.5$~dB (or $<6$~\%). For the analysis
the electric field strength is converted into a power $P=4\pi (r\cdot
E[\mathrm{\mu V\,m^{-1}}])^2/Z_0$, with $log_{10}E[\mathrm{\mu
  V\,m^{-1}}]=E[\mathrm{dB\mu V\,m^{-1}}]/20$ and $Z_0=377\Omega$.

\subsection{\label{subsec:Meas}Measurements}
\begin{figure}
  \centering \includegraphics[width=8cm]{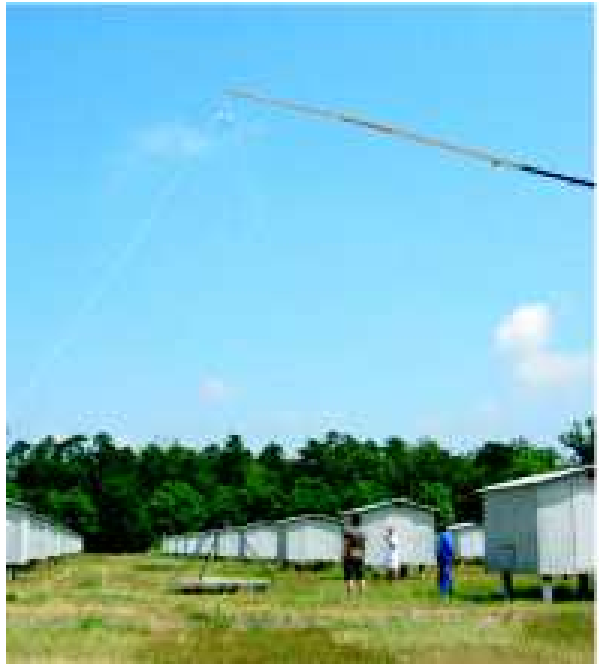}
  \caption{Calibration campaign at the LOPES field.}
  \label{FigPhoto}
\end{figure}

For a calibration campaign, the whole setup of the VSQ is placed at
$r=10.5$~m above each LOPES field antenna, using a movable crane with
a wooden extension (see figure~\ref{FigPhoto}). Fixed in a frame, the
whole setup consists of the reference source VSQ, an external GPS
antenna, and the GPS hand-held unit.  The wooden extension is
necessary to avoid reflections of the radio signal off metal parts of
the crane, which would otherwise lead to differences between
calculated and real received power. For the VSQ setup we use a plummet
with a fixed length and a differential GPS to determine the exact
position of the radio source.
As a calibrated source is used, the power of the radio signal which
arrives at the antenna can be calculated. For the LOPES electronics
we are using an 80~MHz ADC working in the second Nyquist-domain,
digitizing the measured voltages for frequencies from 40--80~MHz.

The calibrated external radio source is emitting at a distance of
$r=10.5$~m and the received signal in the field antenna is transmitted
to the electronics chain via a long coaxial cable. During each
calibration run, artificially triggered data files are stored. Each
file contains N=65536 samples per antenna, at a sampling rate of 
$12.5$~ns this results in $819.2~\mu$s of data. The calibration 
setup lasts for at least two
minutes at a field antenna and uses a trigger period of 6 seconds
(rate 0.17~Hz), which leads to 20 or more stored data files.  This way
of data taking is used to average over small geometric variations in
the setup. On the other hand, influences from wind gusts and fast
changes in environmental conditions can be monitored and affected data
files are not considered for the next analysis steps.

\begin{figure}
  \centering
  \includegraphics[width=8cm]{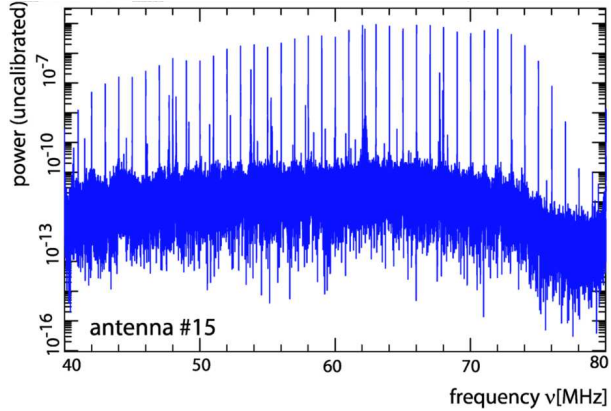}
  \caption{Measured power of one calibration raw data file.
    Clearly visible are the 1 MHz steps of the emission of the
    reference antenna.}\label{FigExamp}
\end{figure}

Each data file is analyzed individually in our software package and
after a Fourier transformation the complex values are used to
calculate the amplitude for a certain frequency.  The resolution in
the frequency domain from 40--80~MHz is determined by the Fourier
transformation to $N/2=65536/2=32768$ bins to
$40~\mathrm{MHz}/32768=1.22~\mathrm{kHz}$.  The characteristic of our
reference source is a comb shaped spectrum with 1, 5, or 10~MHz
spacing. In nearly all cases the 1~MHz spacing is used to get a high
coverage for the LOPES frequency range.  An example is shown in
figure~\ref{FigExamp}.  The amplitude in the frequency domain is
equivalent to the received power, binned with N/2=32768 bins. The
received power $P_\mathrm{M}(\nu)$ for each integer frequency is
determined by summing over 50 bins or 61~kHz around the center of the
peak. The peak value is on average three orders of magnitude higher
than the surrounding noise level and has spreads over roughly 5--10
bins, i.e.\,6--12~kHz.  The noise floor around the frequency peak
contributes with less than 1~\% to the integrated power.  At some
fixed frequencies, not every time, and not in all antennas, man-made
RFI in the same order of magnitude as the received signals affects the
measurement. Therefore a linear interpolation of the received power
replaces these contaminated frequencies.
%

The integrated power $P_\mathrm{M}$ for each integer frequency
averaged over all data files in a calibration step (at least 20 events
in 2 minutes) reflects the overall behavior of the LOPES antenna
system during such a measurement. For each antenna a set of 
measurements$P_\mathrm{M}(\nu)$ exists, which is used to
calculate the ratio $V(\nu)$ of measured to expected power.

\begin{figure}
  \centering \includegraphics[width=8cm]{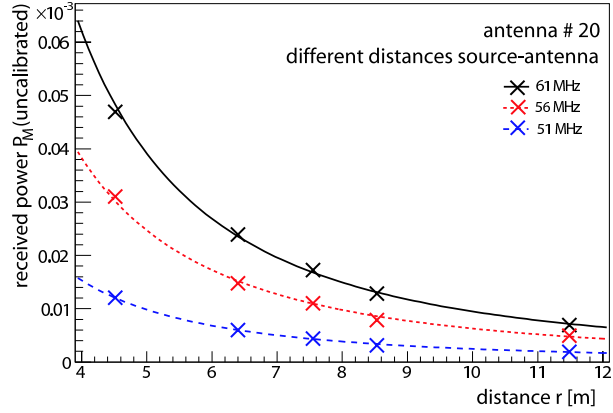}
  \caption{The data points show the received power $P_\mathrm{M}$
    integrated over 61~kHz for various frequencies as a function 
    of $r$, which is the distance source - LOPES antenna. The lines are 
    fits of a $r^{-2}$ dependence.}
  \label{FigRsquare}
\end{figure}

For that purpose one has to be sure that one operates in the far field
region, not to be disturbed by near field effects of the emitter. The
far field approximation ($r_{far} > 2\,D^2\,\nu/c$) should be valid
for $r_{far}>2$~m at 80~MHz and with an antenna aperture $D=1.8$\,. In
the far field approximation the energy density of the electromagnetic
field varies as $r^{-2}$ with the distance $r$.
Figure~\ref{FigRsquare} shows measurements of the received power
$P_\mathrm{M}$ with respect to the distance $r$ between the reference
source and the LOPES antenna. The power $P_\mathrm{M}$ is the
integrated power over 61~kHz around the mean frequency. For
distances of $r=4.5$~m to $r=11.5$~m the far field approximation was
tested. As the fits for the different frequencies (51, 56, and 61~MHz)
are performed with a fixed power index of -2, $P=a\cdot r^{-2}$, the
measurement clearly shows the validation of the far field
approximation. This measurement has also shown that a possible
saturation effect for the low-noise amplifier is unlikely, and that the
calibration acts in the linear working regime. The 12-bit ADC also does not
suffer from a saturation effect, but the closest measurement at 4~m
distance was at the limit of analogue input voltages ($\pm1$~V) for
the ADC. For the calibration campaigns a distance $r$ between field
antenna and reference source of around 10~m was chosen, therefore, we
can exclude saturation effects for the determination of the
amplification factors.

\section{Results}

\subsection{\label{subsec:Ampli}Amplification factors}
\begin{figure}
  \centering
  \includegraphics[width=8cm]{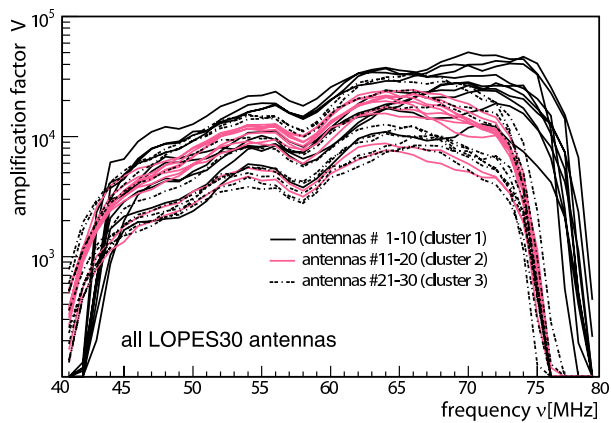}
  \caption{Amplification factors of all LOPES antennas. The
    different colors and line types are used to separate the three
    antenna clusters. Each cluster combines ten antennas with their
    individual electronics.}\label{FigAmpAllAnt}
\end{figure}

The amplification factors $V(\nu)$ for all 30 LOPES antennas are shown
in figure~\ref{FigAmpAllAnt}. The curves for each antenna represent
the mean values of measurements performed over two years.  The frequency
range of the ADC is between 40 and 80~MHz, however, also the filter
characteristics at the upper and lower bound can be seen, which gives
an effective range of approx.\,44 to approx.\,75~MHz. The difference
in the shape of the individual antenna amplification curves is a
result of different types of filters and batches of electronic
components installed for the first 10 antennas compared to the later
installed antenna setups. At the actual configuration of LOPES the
cable length is either 100~m or 180~m, which might also lead to
differences in the amplification factors for different antennas. But
it was found that the total scatter between the amplification factors
(figure~\ref{FigAmpAllAnt}) does not simply relate to the cable
length. Instead, the scatter originates roughly to equal parts from
the adjustments of the operating points of the electronic modules,
peculiarities for certain antenna field positions, and the different
cable lengths.

A common feature of all curves in figure~\ref{FigAmpAllAnt} is a dip
at 58~MHz.  This feature relates to the antenna gain simulation and is
described in more detail in section \ref{subsec:Direct}.

The large overall scatter between the individual antennas of roughly
one order of magnitude obviously necessitates the complete amplitude
calibration. In the analysis of the shower data we use the
amplification factors to correct the raw data in order to get
calibrated power values. Moreover, related systematic effects in the
determination of these amplification factors $V(\nu)$ have been
investigated, which will be described in the following sections.

\subsection{\label{subsec:Stable}Stability of the calibration}
An important issue of the calibration procedure is its stability
over repeated measurements.  Besides uncertainties in the position or
the alignment of the reference source with the field antenna or by
small changes in handling the experimental calibration procedure (see
section~\ref{subsec:Uncer}), there could be large differences between
the measurements of an antenna due to changing environmental
conditions like precipitation, ground humidity, air temperature, etc.
If such differences exist to a greater extent, this would cause
problems in the applicability of the amplification factors.
\begin{figure}
  \centering
  \includegraphics[width=8.8cm]{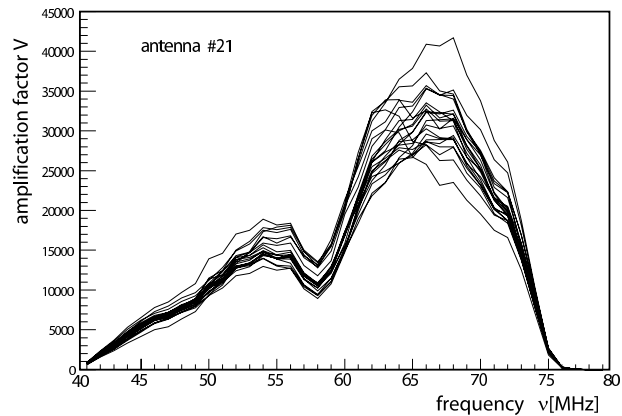}
  \caption{Amplification factor of one LOPES antenna for independent 
    measurements. These measurements were spread over the course of
    nearly two years and include 21 calibration
    campaigns.}\label{FigAmpAnt21all}
\end{figure}
%
To verify the influence of such changing environmental conditions we
performed a series of measurements over roughly two years. In these
campaigns one field antenna (antenna \#21) was measured each time and
therefore under different environmental conditions. The
resulting amplification factors for this antenna are displayed in
figure~\ref{FigAmpAnt21all}.  The measurements include weather
conditions occurring during the year in Karlsruhe, except extremes
like snow and thunderstorm. The temperature ranges from around
$\sim-5~^\circ$C up to $\sim35~^\circ$C, the soil change from dry in
the surface layer (0--60~cm) to a high ground humidity after days of
rain, and the ground vegetation varies from recently mowed to a
flowery meadow. The variation found is a measure of the systematic
uncertainties of the calibration procedure or, on the other hand,
reflects the accuracy of this calibration method. In the effective
frequency range of the LOPES band pass filter from 44 to 75~MHz an
average scatter of about 9~\% was found.

It is obvious that the variation from antenna to antenna
(fig.~\ref{FigAmpAllAnt}) is much larger than the variation due to
changing environmental conditions or conditions of the calibration
procedure (note the logarithmic scale in fig.~\ref{FigAmpAllAnt}
compared to the linear scale in fig.~\ref{FigAmpAnt21all}).

\begin{figure}
  \centering
  \includegraphics[width=8.8cm]{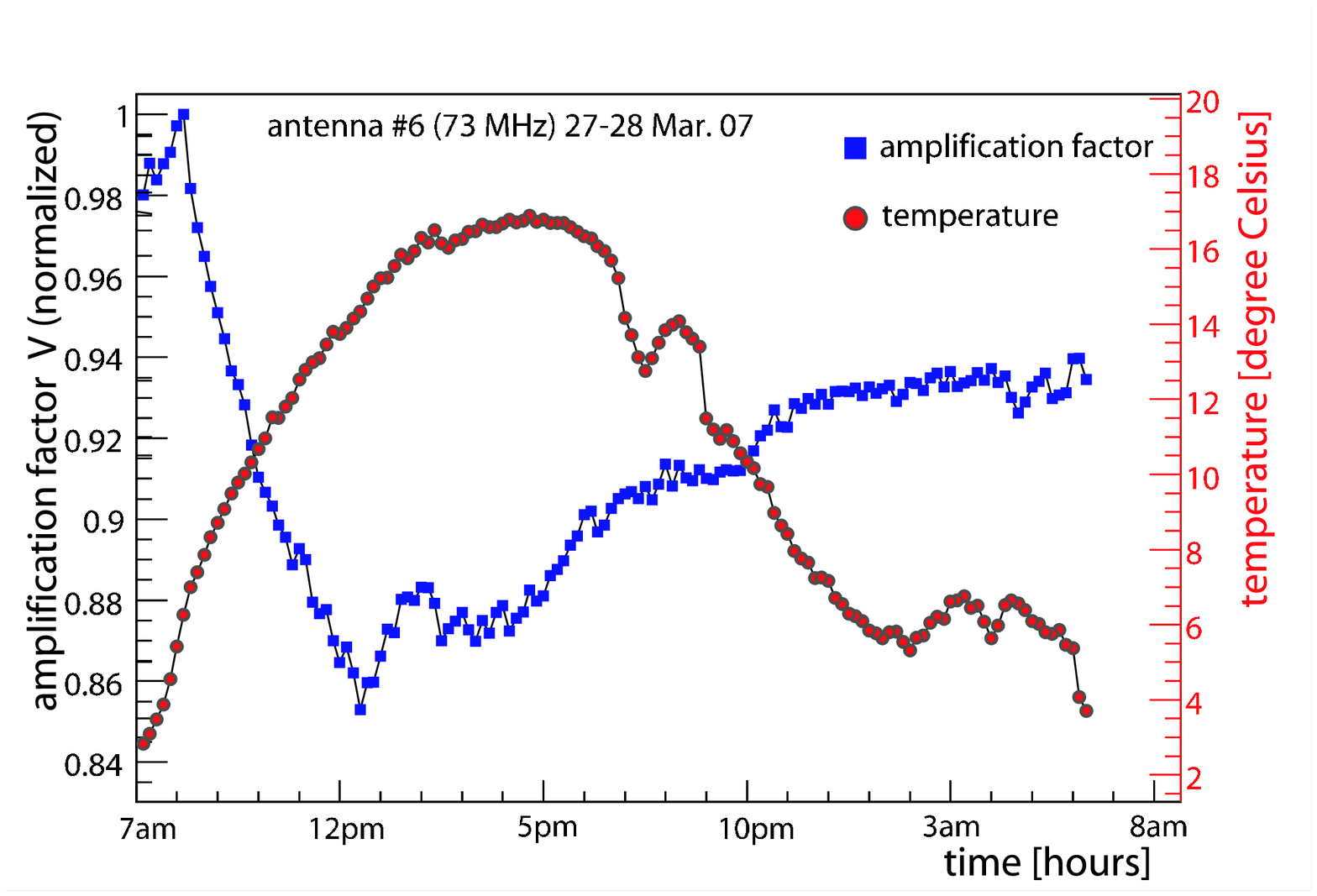}
  \caption{Amplification factor of antenna 6 at 73~MHz (left y-axis) 
    and temperature 2~m above ground (right y-axis) plotted against
    time. The antenna was monitored over a period of 24 hours with
    fixed setup started at 7~a.m. and the highest value was used to
    normalize the amplification factor $V$. The maximum drop for $V$
    is obtained at 12~p.m. and add up to around 15\%.}
  \label{FigTempAnt6}
\end{figure}

There are many possible sources for the systematic uncertainties and a
significant reduction of the observed variation requires deeper
investigations. One possible source of the variations is thought to be
the LNA temperature. But the temperature of the low noise amplifier at
the antenna and at one end of the long coaxial cable is not directly
available. The remaining part of the cable and the LOPES electronics
is either underground or inside an air-conditioned housing. However,
variations with the air temperature can be used as a first order
approximation. We use the air temperature at two meters height
above ground, which is provided by the KASCADE experiment.  With 24 hours
continuous monitoring using the VSQ at a fixed position in a distance
of 16.5~m and with a zenith angle of $87^\circ$, we covered an air
temperature range of $3~^\circ$C -- $28~^\circ$C. The data taking is
similar to the calibration campaigns, except that here the KASCADE
trigger is used, with an average rate of 0.04~Hz.  For each ten
minutes an averaged amplification factor $V(\nu)$ is calculated.
Figure~\ref{FigTempAnt6} displays the average amplification factor
$V(73~\mathrm{MHz})$ (squares, left y-axis) and the air temperature in
degrees at 2~m height (circles, right y-axis) over time. To show 
the relative change of the average amplification factor, the highest
value at the day was used for normalization. The daily temperature
modulation is well pronounced and shows a minimum in the early
morning. The highest temperature was recorded in the early afternoon.
As the performance of electronic devices changes with temperature, the
LNA circuit connected directly to the antenna at the field has a
better performance at low temperature values.
%

From figure~\ref{FigTempAnt6} one derives that in general the
amplification factor decreases to lower values with increasing air
temperature. The maximum drop add up to around 15\% for the
amplification factor at this day and is obtained at 12~pm. For
the covered air temperature range one can derive that ten
degree difference in temperature causes a change of about 10~\% in the
amplification factor. This implies that a first order correction of
this effect is possible using the air temperature.  The uncertainty in
temperature stability of about 6~\% of the reference signal generator
is not corrected for. This value is considered for the temperature
range $10~^\circ$C -- $30~^\circ$C. The change in temperature is
somewhat higher than this methodical uncertainty, but can not be
disentangled.  Nevertheless, the LNA temperature should be measured
directly to fully verify the temperature dependence of the LOPES
system.

\subsection{\label{subsec:Polar}Cross-check of the polarization sensitivity}
\begin{figure}[t]
  \centering
  \includegraphics[width=9cm]{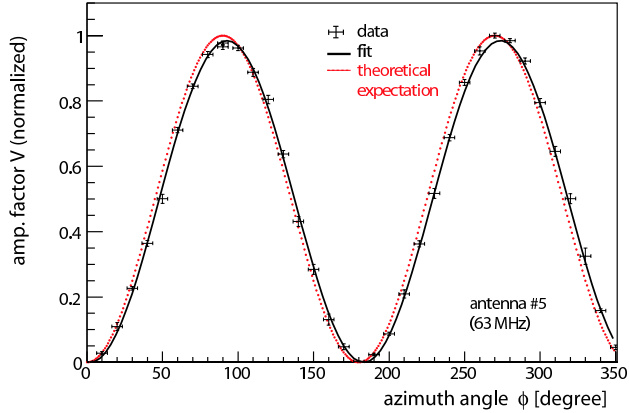}
    \caption{Relative amplification factors V at 63~MHz for the 
      dual-polarized antenna 5, but with varying the azimuth angle
      $\phi$ and normalized to the amplification factor at
      $\phi=270^\circ$. The position of the source was vertical above
      the LOPES antenna. The fitted function
      $V(\phi)=a\cdot\cos^2(\phi+\alpha-\phi_{\mathrm{Ant}})$ is
      indicated by a black line and the theoretical expectation
      $V(\phi)=\cos^2(\phi-\phi_{\mathrm{Ant}})$ by a red dashed line,
      both are normalized to the value at $\phi=270^\circ$).}
    \label{FigPol}
\end{figure}

Since the end of 2006 LOPES is operating some antennas in a dual-polarized
configuration. In particular, for such a configuration a cross-check
of the polarization sensitivity has to be performed during the
calibration campaigns. Figure~\ref{FigPol} shows the relative
amplification factors for the east-west oriented antenna \#5 measured
in one campaign where the azimuth angle $\phi$ was varied in steps of
10 degrees by rotating the VSQ reference source above the field
antenna ($\phi_{\mathrm{Ant}}=90^\circ$, equal to east-west
orientation). The amplification factor $V$ at $\phi=90^\circ$ was
used for normalization. From formula~\ref{Formel_V} we know that the
received power changes with $\cos^2\beta$, with respect to the linear
polarized LOPES antenna. The angle $\beta$ is defined as the angle
between the axes of VSQ and field antenna. The azimuth angle $\phi$ is
related with the angle $\beta$ by $\phi=\beta-\phi_{\mathrm{Ant}}$.

The result obtained in figure~\ref{FigPol} prove the expected
$\cos^2\beta$ polarization sensitivity of the LOPES antenna, here
exemplarily shown at 63~MHz. The fitted function
$V(\phi)=a\cdot\cos^2(\phi+\alpha+\phi_{\mathrm{Ant}})$ uses two free
parameter $a=0.98\pm 0.01$ as scaling parameter and
$\alpha=-3.0^\circ\pm 0.6^\circ$ as angle offset to describe the
measurements. The theoretical expectation is based on a
$V(\phi)=cos^2(\phi+\phi_{\mathrm{Ant}})$ with no free parameters.
Both curves are very close to each other and show the achieved
accuracy for aligning the polarization axes from VSQ and field
antenna. With a systematic offset of $\alpha=-3.0^\circ$ for this
measurement we derive a systematic uncertainty of
$\sigma_\mathrm{\beta}=7^\circ$.  However, in the case of strong
disturbing wind the alignment might be worse and can result in a
higher loss in the received power.

\begin{figure}
  \centering
  \includegraphics[width=9cm]{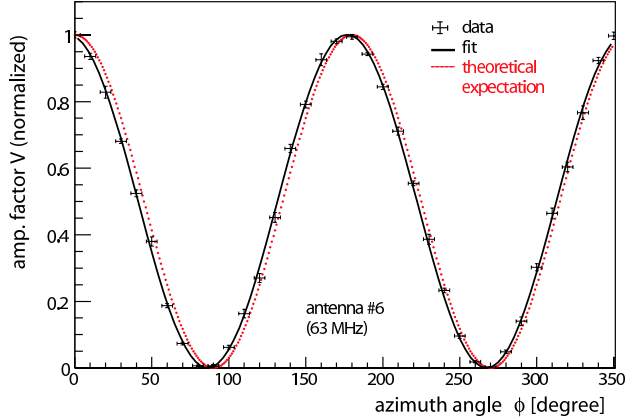}
  \caption{Same as Fig.~\ref{FigPol}, but for the perpendicular 
    oriented antenna 6, i.e. an antenna sensitive to the north-south
    polarization of the signal. Here the fitted function and
    theoretical expectation are normalized to the value at
    $\phi=180^\circ$.}
  \label{FigPolNS}
\end{figure}

%
Figure~\ref{FigPolNS} shows results of the same measurements at the
same antenna station, but the data are analyzed for the north-south
oriented antenna \#6, i.e.  $\phi_{\mathrm{Ant}}=0^\circ$.  Together
with figure~\ref{FigPol} the two plots show the capability and
applicability of calibration and event data taking with LOPES in a
dual-polarized antenna mode. The measured power ratio of antenna \#6
to antenna \#5 is less then 2~\% for the relevant frequency range.
Despite this independent calibration there remains the possibility of
cross talk between both channels as their low noise amplifiers are
mounted in the same box.
%

\subsection{\label{subsec:Uncer}Uncertainties}
In previous sections systematic effects of the calibration source, the
temperature dependence, and the polarization sensitivity were
described. In this section we will discuss in detail the different
sources of uncertainties and summarize them. The statistical
uncertainty of the calibration procedure for the amplification factors
$V(\nu)$ in the effective frequency range 44--75~MHz are in most cases
negligible ($\mathrm{stat_{calib}}=1.5$~\%) due to the 20
'trigger'-measurements per individual calibration configuration.

The total systematic uncertainty for one calibration campaign is
estimated by a combination of the individual sources. These sources
are identified and the resulting standard deviations estimated for:
\begin{description}
  
\item[r:] The uncertainty in estimating the distance between reference
  source and antenna is estimated to $\sigma_{r}=0.25$~m.
  
\item[\boldmath{$\nu$}:] The deviation in frequency is given by the
  resolution of the power spectrum after fast Fourier transformation
  and results to $\sigma_{\nu}=600$~Hz.
  
\item[\boldmath{$G_\mathrm{t}P_\mathrm{t}$}:] The information from the
  data sheet of the reference antenna allows only to estimate the
  uncertainty of the product of gain and power. There is a variation
  of the output power with changing temperature in the order of
  $\sigma_{G_\mathrm{t}P_\mathrm{t}}/G_\mathrm{t}P_\mathrm{t}=12$\%.
  This does not include the systematic uncertainty for the calibration
  of the reference source itself, which is $2.5$\,dB for the electric
  field, respectively $\mathrm{sys_{reference}}\approx 67$~\% for the
  emitted power.
  
\item[\boldmath{$G_{\mathrm{r}}$}:] The simulated antenna gain. This
  is second largest source of uncertainty and the largest for the
  calibration procedure itself and is estimated to be
  $\sigma_{G_{\mathrm{r}}}/G_{\mathrm{r}}\approx 15$~\%, but can be
  even larger at the expected resonant frequency; see next section.
  
\item[\boldmath{$P_\mathrm{M}$}:] The uncertainty of the measured
  power is given by the data acquisition system and read-out process
  of the power values and estimated to be
  $\sigma_{\mathrm{P_\mathrm{M}}}/{P_\mathrm{M}}\approx5$~\%.
  
\item[\boldmath{$\beta$}:] The uncertainty for the angle between the
  polarization axis of the reference source and the axis of the LOPES
  field antenna is estimated to $\sigma_\mathrm{\beta}=7^\circ$, which
  results in a maximum loss of 2~\% for the emitted power.
  
\item[Environmental effects:] Due to the fact that the LOPES antenna
  uses the ground as a reflector, humidity at the ground can influence
  the values, despite the metal reflector below the antennas. An
  estimation of this uncertainty is derived from performing
  measurement campaigns over two years (figure  ~\ref{FigAmpAnt21all}) 
  and results to $\approx9$~\%.  In this uncertainty, effects from the
  antenna coupling, varying ground conditions, and a temperature 
  changes of the signal generator output are included.

\end{description}

From these systematic uncertainties an overall uncertainty for the
amplitude calibration of the LOPES antenna array can be calculated to
$\sigma_V/V=0.70$. This includes the statistical uncertainty
$\mathrm{stat_{calib}}$ and all other uncertainties described above.
As these uncertainties are of different kind a separation in three
main groups can be done:
\begin{eqnarray}
\left(\frac{\sigma_V}{V}\right)^2 &=& \left(\mathrm{stat_{calib}}\right)^2 + 
                                      \left(\mathrm{sys_{calib}}\right)^2 +
                                      \left(\mathrm{sys_{reference}}\right)^2\nonumber\\
&=& (0.015)^2 + (0.205)^2 +(0.67)^2
\end{eqnarray}
Ignoring the calibration uncertainty given by the commercial radio
source with \mbox{$\mathrm{sys_{reference}}=67\%$} the listed
systematic uncertainties sum up to $\mathrm{sys_{calib}}=0.205$. Here,
the antenna simulation and the environmental effects give the largest
contributions. By more detailed studies of the antenna directivity the
uncertainty $\mathrm{sys_{calib}}$ might be decreased
(section~\ref{subsec:Direct}).

As the weather and environmental effects are difficult to quantify, a
correction of a correlation of the system performance with the
temperature (as described in section~\ref{subsec:Stable}) is not yet
performed for the amplification factors. On the other hand it will
improve the uncertainty for the environmental effects only, which
contributes only with $\approx 9$~\%.

The dominating factor for the total uncertainty is given by the
calibration accuracy of the VSQ1000 radio source, provided by the
manufacturer. Using another reference source would lead to a
different, maybe smaller, total systematic uncertainty.

\subsection{\label{subsec:Direct}Cross-check of the antenna directivity}
\begin{figure}
  \centering
  \includegraphics[width=8cm]{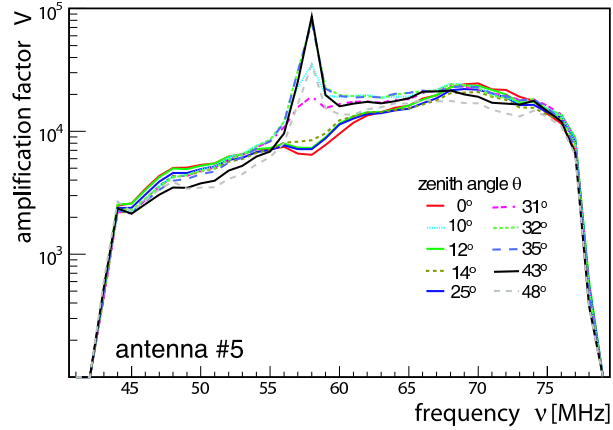}
    \caption{Amplification factor distributions for one antenna, 
      but calibrated with the reference source located at different
      distances, i.e. emission in different inclination angles with
      respect to the LOPES antenna.}
    \label{FigAngle}
\end{figure}

The described calibration procedure also allows us to check at least
partly the simulation results (section~\ref{subsec:Simu}) of the
antenna gain pattern.  The prediction of a pronounced resonance at
58~MHz caused a dip, visible in figure~\ref{FigAmpAnt21all}, and is
introduced by the factor $G_{\mathrm {r}}$ in the calculation of the
amplification factors.  By moving the source away from the zenith, but
tilting the source accordingly to keep the emission and the
polarization angles constant, one is able to check the zenithal
dependence of the antenna gain. For higher inclinations the
amplification factors, i.e. the sensitivity of the antenna, decreases.
Figure~\ref{FigAngle} shows the amplification factors for one antenna
measured at the same campaign but the source located at different
zenith angles ($\theta \ne0^{\circ}$). Due to the fact that in the
calculation of the amplification factor V (see eqn.~\ref{Formel_V})
the simulated antenna pattern $G_{\mathrm {r}}$ is included, the
obtained distribution of the amplification factors in
Fig.~\ref{FigAngle} should be always the same in the range of the
uncertainty.  This is true for a large range of the frequency band,
but not around the frequency of 58~MHz. At this frequency, a resonance
effect of the metal ground plate is predicted by the simulations (see
figure~\ref{FigSimuXZ}).

%
\begin{figure}
  \centering
  \includegraphics[width=8cm]{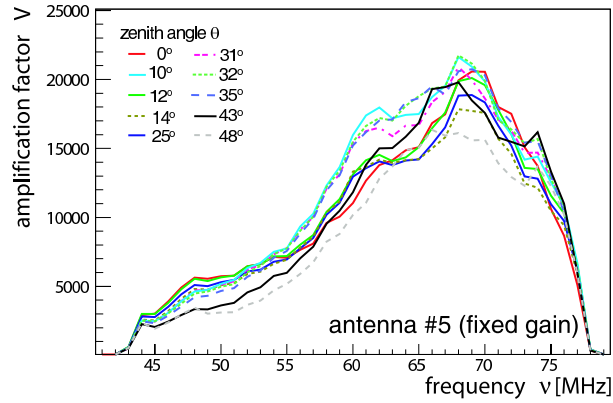}
    \caption{Same as Fig.~\ref{FigAngle}, but this time the 
      amplification factor V was calculated using a fixed antenna gain
      $G_\mathrm{r}=4.0$ and plotted with a linear scale for the
      y-axis.}
    \label{FigAngleFix}
\end{figure}
%
Figure~\ref{FigAngleFix} shows results of the same measurement, but
now V calculated with a fixed antenna gain.  Choosing for
$G_\mathrm{r}(\theta,\phi,\nu)=4.0$ (value for vertical source
position at $50$~MHz) the antenna pattern corresponds to an isotropic
radiator.  If the resonance effect would be present as predicted, a
clear peak for vertical positions and a large dip for $\theta =
45^\circ$ should be visible at 58~MHz. As this is not the case we
conclude that either the calibration procedure smears out all the
effects, which is implausible with the obtained total systematic
uncertainty of the calibration, or the antenna gain simulation
overestimates the effect of the ground plate.  In addition, the
results shown in Fig.~\ref{FigAngleFix} suggest that the resonance is
at 62~MHz rather than at 58~MHz and not as sharp as expected by the
simulations.

\section{Summary}
In this work we studied a method for an absolute amplitude calibration
of a digital radio antenna array using a calibrated external reference
source. By application of the procedure to the LOPES antenna array we
could show that it is possible to derive reliable, frequency dependent
amplification factors allowing us to measure the absolute electric
field strength emitted by cosmic ray air showers.

The absolute amplitude calibration was successful and detailed
investigations of possible systematic uncertainties lead to a total
uncertainty of $\sigma_V/V=70$~\%. This total uncertainty also
includes e.g. environmental effects, like those caused by different
weather conditions present over nearly two years of calibration
campaigns ($\mathrm{sys_{calib}}=20.5$~\%), and a systematic
uncertainty from the used reference source
($\mathrm{sys_{reference}}=67$~\%).

Electronic modules are temperature dependent and we have shown that
there is a relation between air temperature and amplification factor
$V(\nu)$ for the LOPES antenna system. A more precise correlation
analysis and following correction can improve the overall precision
for measuring electric field strengths.

The systematic uncertainty of the calibration of the reference radio
source itself contributes with $\approx67$~\% to the total
uncertainty.  Using another, more precisely calibrated reference radio
source would improve the accuracy.

The measurements at a dual-polarized antenna setup with two low noise
amplifiers and cables next to each other indicate that a calibration
of such a system is possible and reliable. There seems to be no
significant cross-talk between the channels. The electric field
strength for such a configuration can be measured with the same
accuracy as for a single polarized antenna.

The antenna gain simulation contributes to a large amount to the total
uncertainty.  The investigation of the predicted resonance at 58~MHz
indicate a much weaker influence of the metal pedestal than expected.
The dip feature (figure~\ref{FigAmpAnt21all}) at the resonance
frequency is introduced by the calculation of the amplification
factors and does not appear during the investigation using a spherical
antenna gain pattern. The present investigations have shown that the
simulations used for the antenna gain have to be checked in greater
detail to reveal the influence of the pedestal. Measurements with an
increased coverage of the directional pattern can help to improve the
simulations.

An important conclusion of this work is that the discussed strategy of
calibration can be adapted for future radio antenna arrays measuring
cosmic ray air showers. Especially at locations with much lower RFI an
astronomical source, e.g. the galactic background radiation, can be
used to cross-check the proposed amplitude calibration of a radio
antenna system.

\section*{Acknowledgments}
  The authors would like to thank the technical staff of the
  Forschungszentrum Karlsruhe for their enthusiastic help during the
  calibration campaigns in the field, in particular Mr. Edgar
  F\"ussler, the crane driver, who always had fun and patience for the
  positioning of the reference source. Sincere thanks to the entire
  LOPES collaboration for providing the working environment for these
  studies.

\end{document}